\def\vec#1{{\bf#1}}

\documentclass[12pt,preprint]{aastex}

\shorttitle{Sources of the Slow Wind}
\shortauthors{Antiochos et al.}

\begin{document}

\title{A Model for the Sources of the Slow Solar Wind}


\author{S.\ K.\ Antiochos}
\affil{NASA Goddard Space Flight Center, Greenbelt, MD, 20771}
\email{spiro.antiochos@nasa.gov}

\and

\author{Z. Miki\'c, V.~S.\ Titov, R. Lionello, J.~A.\ Linker}
\affil{Predictive Science, Inc., San Diego, CA 92121}


\begin{abstract}

  Models for the origin of the slow solar wind must account for two
  seemingly-contradictory observations: The slow wind has the
  composition of the closed-field corona, implying that it originates
  from the continuous opening and closing of flux at the boundary
  between open and closed field. On the other hand, the slow wind also
  has large angular width, up to $\sim 60^\circ$, suggesting that its
  source extends far from the open-closed boundary. We propose a model
  that can explain both observations. The key idea is that the source
  of the slow wind at the Sun is a network of narrow (possibly
  singular) open-field corridors that map to a web of separatrices and
  quasi-separatrix layers in the heliosphere. We compute analytically
  the topology of an open-field corridor and show that it produces a
  quasi-separatrix layer in the heliosphere that extends to angles far
  from the heliospheric current sheet. We then use an MHD code and
  MDI/SOHO observations of the photospheric magnetic field to
  calculate numerically, with high spatial resolution, the
  quasi-steady solar wind and magnetic field for a time period
  preceding the August 1, 2008 total solar eclipse. Our numerical
  results imply that, at least for this time period, a web of
  separatrices (which we term an S-web) forms with sufficient density
  and extent in the heliosphere to account for the observed properties
  of the slow wind. We discuss the implications of our S-web model for
  the structure and dynamics of the corona and heliosphere, and
  propose further tests of the model.

\end{abstract}


\keywords{Sun: magnetic field --- Sun: corona --- Sun: solar wind}


\section{Introduction}

Decades of in situ measurements of the heliosphere have firmly
established that the Sun's wind consists of two distinct types:
``fast'' and ``slow''. In terms of its origins at the Sun, the best
understood is the fast wind, which typically exhibits speeds in excess
of 600 km/s at 1 AU and beyond \citep[e.g.,][]{mccomas08}. The fast
wind is measured to be approximately steady, except for some
Alfv\'enic turbulence \citep[e.g.,][]{bame77, bruno05}. This wind is known to
originate from coronal holes, regions that appear dark in XUV and
X-ray images, due to a plasma density that is substantially lower
($<50\%$) than in surrounding coronal regions \citep{zirker77}. As
implied by eclipse and coronagraph images, the magnetic field in
coronal holes is open---appearing mainly radial and stretching out
without end---whereas the field in the surrounding regions is closed,
looping back down to the photosphere. Hence, the fast wind corresponds
to the steady wind predicted by Parker in his classic work
\citep{parker58, parker63}.

The slow wind, however, is much less understood. In particular, its
origin at the Sun has long been one of the major unsolved problems in
solar/heliospheric physics. This wind has a number of observed
features that distinguish it physically from the fast wind. First, its
speeds are typically slower, $< 500\,{\rm km/s}$.  More important, the
slow wind appears to be inherently non-steady when compared to the
fast wind \citep[e.g.,][]{bame77, schwenn90, gosling97, mccomas00}. It
exhibits strong and continuous variability in both plasma (for
example, speed and composition) and magnetic field properties;
variability that cannot be described as simply Alfv\'enic disturbances
superimposed on a steady background \citep{zurbuchen06,
  bruno05}. Finally, its location in the heliosphere is distinct; it
is generally found surrounding the heliospheric current sheet (HCS)
\citep[e.g.,][]{burlaga02}. A key point is that the HCS is always
embedded inside slow wind, never fast. From the presently available
spacecraft observations, it is not possible to rule out the
possibility that slow wind also occurs in locations unconnected to the
HCS, in other words, that there are pockets of slow wind with no
embedded HCS and surrounded completely by fast wind. However, the
present data are certainly consistent with the picture that, at least,
during solar minimum when the corona-wind mapping can be determined
with some accuracy, all slow wind originates from a band that
encompasses the HCS, so that the mapping of the slow wind down to the
Sun appears to connect to or near the helmet streamer belt
\citep[e.g.,][]{gosling97, zhao09}.

Another key feature of the slow wind is its latitudinal extent, which
typically ranges from $40^\circ$--$60^\circ$ near solar minimum, a
time when it is easiest to distinguish the sources of fast and slow
wind. Within this broad region of slow wind the actual HCS, across
which the magnetic field changes direction, is very narrow.  As for
any current sheet, one can identify in the heliospheric data a scale
over which the field becomes small and the plasma beta, defined as the
ratio of the gas pressure $P_g$ to the magnetic pressure $B^2/8\pi$,
becomes large. This region is termed the plasma sheet and is usually
of the order of a few degrees in angular width
\citep[e.g.,][]{winterhalter94, bavassano97, wang00, crooker04}. It is
important to note that the HCS is often not symmetrically located
within the broad band of slow wind, but is often found nearer to one
edge of the slow wind region \citep{burlaga02}. It is also important
to note that the field almost never vanishes at the HCS, as would be
expected for a true steady-state. This observation implies that, at
least, the wind near the HCS must be continuously dynamic.

The final and most critical feature of the slow wind that
distinguishes it from the fast is the plasma composition
\citep{geiss95, vonsteiger95}. It is well-known that in the closed
field corona, the ratio of the abundances of elements with low first
ionization potential (FIP), such as Mg and Fe, to those with high FIP,
such as C and Ne, is a factor 4 or so higher than in the photosphere
\citep[e.g.,][]{meyer85, feldman03}. This so-called FIP effect is not
seen in the fast wind, which has abundances similar to those of the
photosphere; but, it is present in the slow wind, which has abundances
similar to that of the closed corona \citep{gosling97, zurbuchen06,
  zurbuchen07}. 

Along with the difference in elemental abundances, the slow and fast
wind also exhibit clear differences in their ion charge state
abundances, for example, the ratio of ${\rm O}^7/{\rm O}^6$. This
ratio can be used to determine the ``freeze-in'' temperature of the
ion charge states at the source of the wind. Close to the Sun where
the time scales for ionization and recombination are much shorter than
the plasma's expansion time-scales, the ion charge states are
approximately in ionization equilibrium with the local electron
temperature. As the solar wind plasma expands outward, however, the
electron density drops rapidly and the recombination time scales
become so large that the ionic charge states stop changing,
freezing-in the electron temperature at this point. The freeze-in
radius varies for the different ions, but is typically 1 - 3
R$_\odot$. The data show that the slow wind has a higher freeze-in
temperature ($\ge 1.5 \times 10^6\,{\rm K}$) than the fast wind ($\le
1.2 \times 10^6\,{\rm K}$) \citep{vonsteiger97, vonsteiger01,
  zurbuchen99, zurbuchen02}. Note, however, that this freeze-in
temperature corresponds only to the electron temperature in the low
corona. The proton and ion temperatures measured in situ and in
coronal holes by UVCS, for example, \citep[e.g.,][]{kohl06} show the
opposite trend in that the ion temperatures are substantially higher
in the fast wind than in the slow \citep{marsch06}. The origin of
these differences in the ion temperatures between the two winds is
still not clear, but in any case, both the ion and freeze-in
temperatures suggest that the sources of the two winds near the Sun
are physically different.

The elemental abundances track very well the ionic abundances,
indicating that there is a consistent compositional distinction
between the two winds. Furthermore, the two winds have markedly
different temporal variability in elemental and ionic composition. The
fast exhibits an approximately constant composition; whereas the slow
exhibits large and continuous variability, so that its elemental
composition varies from coronal to near photospheric. The
composition results suggest that the fast wind has a unique origin,
presumably in coronal holes, but that the slow wind originates from a
mixture of sources.

In fact, Zurbuchen and coworkers have argued that the compositional
differences, rather than the speed, are what truly distinguish the two
winds, because it is possible to find solar wind whose composition and
constancy match that of the ``fast wind,'' but that has relatively
slow speed, $< 500\,{\rm km/s}$ \citep{zhao09}.  Note also that, as
determined by the composition measurements \citep{zurbuchen99}, the
boundary between the slow and fast wind in the heliosphere is sharp,
of order a few degrees in angular extent, much smaller than the
angular width of the slow wind region, but comparable to that of the
plasma sheet. An important point is that the observed sharpness of the
composition transition is not merely a dynamical effect, because it
does not depend on whether the stream-stream transition is fast to
slow or slow to fast \citep{geiss95, zurbuchen07}. We conclude,
therefore, that the fast and slow winds are far more appropriately
described as the steady and unsteady winds, and that the boundary
layer between the two winds is much narrower than the width of either
wind.

Since the differences in plasma composition of the two winds must be
due to differences in their origins at the Sun, the composition data
place severe constraints on the possible sources of the slow wind. In
particular, the data imply that the slow wind originates in the
dynamic opening of closed magnetic flux, which releases
closed-corona plasma into the wind. Such a process would also
naturally explain the difference in variability between the fast and
slow wind.

It should be emphasized, however, that this constraint on the slow
wind's origin is not universally accepted. Several authors have argued
that the slow wind originates from open-field coronal holes, just like
the fast wind, but from the {\it edges} of the holes, where the field
expands super-radially as it extends from the photosphere out to the
heliosphere \citep[e.g.,][]{kovalenko81, wang91, cranmer05, cranmer07,
  wang09}. The hypothesis is that a large expansion factor can both
slow down the wind by affecting the location of wave energy deposition
in coronal flux tubes, and change the plasma composition by the FIP
mechanism proposed by \citet{laming04}. Note that in the expansion
factor model, as in all steady state wind solutions, the properties of
the wind in a given flux tube are determined uniquely, in most cases,
by the flux tube geometry and the forms of the heating and momentum
deposition \citep{cranmer07}. Of course the detailed forms of the
heating and momentum deposition will depend on the flux tube geometry,
and may depend on other factors, as well, but the dependence on these
other factors cannot be dominant; otherwise the calculated wind speed
would not be well correlated with expansion factor. In other words,
two flux tubes on the Sun with identical geometry should have similar
heating/momentum deposition and end up with the same wind
properties. Therefore, the steady-state models inherently predict a
tight correlation between speed and composition
\citep[e.g.,][]{cranmer07}.

The problem, however, is that observations indicate that wind speed is
not tightly correlated with composition. The wind from small
equatorial coronal holes with a large expansion factor is indeed slow,
with speeds $< 500\,{\rm km/s}$, in good agreement with the
predictions of the expansion factor models.  But this wind has
photospheric FIP ratios, so it is still considered to be ``fast wind''
\citep{zhao09}. Furthermore, this not-so-fast wind has the temporal
quasi-steadiness of the fast wind, rather than the quasi-chaotic time
variation of the slow wind.

We conclude, therefore, that the most likely source for the true slow
wind, that with FIP-enhanced coronal composition, is the closed-field
corona. In this case, the process that releases the coronal plasma to
the wind must be either the opening of closed flux or interchange
reconnection between open and closed magnetic field lines. This latter
process is the underlying mechanism invoked by Fisk and co-workers
\citep{fisk98, fisk03, fisk09} in their model for the heliospheric
field. These authors argue that open flux can diffuse freely
throughout the solar surface, even deep inside the helmet streamer
region. If so, then the interchange reconnection between open and
closed magnetic field lines would naturally account for both the
composition and geometrical properties of the slow wind. The
difficulty with this model is that it has not been demonstrated that
such open flux diffusion can actually occur. In fact, detailed MHD
simulations indicate that it is difficult to bring open fields into
closed-field regions without having them close down
\citep{edmondson10, linker10}. The simulation results are in agreement
with \citet{ska07}, who argued that, for the low-beta corona, basic
MHD force balance forbids the presence of open flux deep inside the
closed helmet streamer region.

Within the context of MHD models, the most likely location for the
release of closed-field plasma is from the tops of helmet streamers
(the Y-point at the bottom of the HCS), where the balance between gas
pressure and magnetic pressure is most sensitive to perturbations. A
number of authors have argued that streamer tops are unstable and
should undergo continual opening and closing as a result of thermal
instability \citep{suess96, endeve04, rappazzo05}.  Even if streamer
tops are stable, it seems inevitable that the constant emergence and
disappearance of photospheric flux and the constant motions of the
photospheric would force them to be continuously
evolving. Furthermore, coronagraph observations often show the
ejection of ``blobs'' from the tops of streamers and into the HCS
\citep{sheeley97}.

Although this streamer top model seems promising in that it naturally
explains both the composition and variability, it has difficulty in
accounting for the large angular widths of the slow wind. One would
expect the instabilities to be confined to the high-plasma beta region
about the current sheet. In fact, the plasma emanating from the
streamer tops, the so-called stalks, is observed to be only $\sim
3^{\circ}$--$6^{\circ}$ wide, which agrees well with the plasma sheet
width in the heliosphere \citep{bavassano97, wang00}.  Even if the
plasma sheet width were to be widened by the Kelvin-Helmholtz
instability \citep[e.g.,][]{einaudi99}, there would not be enough mass
flux from the narrow region at the streamer tops to account for the
slow wind. The streamer-top models can account for a thin band of slow
wind around the HCS, but it seems unlikely that this is the origin of
the bulk of the slow wind, which can extend as far as $30^{\circ}$ in
latitude from the HCS.

In order to be compatible with the in situ data, we require some
process that releases closed-field plasma onto open field lines that,
in the heliosphere, can be far from the HCS. This requirement seems
impossible to satisfy, because the plasma release must occur at the
boundary between the open and closed field in the corona, which maps
directly to the HCS. We describe below, however, a magnetic topology
that resolves this slow wind paradox: the flux associated with an
open-field corridor can be simultaneously near to and far from the
open-closed boundary!

\section{The Topology of an Open-Field Corridor}

Figure~1 illustrates the magnetic connectivity from the photosphere to
the heliosphere that results from an open-field corridor.  The dark
yellow inner sphere in the figure represents the photosphere, while
the light yellow, semi-transparent one represents an arbitrary radial
surface in the open-field heliosphere, say at $5R_\odot$ The green
line on the photosphere marks the boundary between open (gray) and
closed (yellow) field regions, which is mapped by the magnetic field
(red lines) to the HCS (thick green line) at the $5R_\odot$
surface. The green line at the HCS is also the polarity inversion line
at this surface. Note that the four points, a, b, c, and d, which are
meant to represent the end-points of the corridor at the Sun, map
sequentially to the corresponding points a$^\prime$, b$^\prime$,
c$^\prime$, and d$^\prime$ along the HCS.
 
The open field pattern at the photosphere of Fig.~1 consists of a
large polar coronal hole and, as is often seen, a smaller low-latitude
hole. In recent work, we argued that if the two holes are in the same
photospheric polarity region, then by our uniqueness conjecture the
holes must be connected by an open field corridor, as illustrated
above \citep{ska07}. It is evident from the figure that the flux in
the corridor maps on the heliospheric surface to a thin arc (light
gray band), bounded at both ends by the HCS. The flux between the arc
and the HCS maps to the low-latitude extension while the flux outside
the arc maps to the main part of the polar coronal hole.  The corridor
and its associated arc are the footprints of two quasi-separatrix
layers \citep[QSLs, e.g.,][]{priest95,demoulin96} that combine
into a hyperbolic flux tube, as has been described in detail by
\citet{titov02,titov08} for the case of closed magnetic
configurations.  In contrast, the HCS is a true separatrix.

The key point for understanding the origin of the slow wind is that,
just like the HCS, the QSL arc in the heliosphere can also be a source
region for slow wind. If the open-field corridor at the Sun is
sufficiently narrow, then the continual evolution of the photosphere,
driven by the ever-present supergranular flow and flux
emergence/submergence in particular, will continually change the exact
location of this corridor. But, by the uniqueness conjecture
\citep{ska07}, the corridor is a topologically robust feature, similar
to a null-point, and must be present on the photosphere as long as the
low-latitude coronal hole extension is present. Its location and
shape, however, will vary in response to local photospheric
changes. These variations require field line opening/closing and
interchange reconnection, thereby releasing closed-field plasma all
along the QSL arc in the heliosphere. Therefore, if the QSL arc
extends to high latitudes, this will naturally produce slow wind with
an extent far from the HCS.

To determine whether the QSL resulting from an open field corridor
does, indeed, reach high heliospheric latitudes, we have calculated an
example of a field such as that of Fig.~1 using the source surface
model \citep{altschuler69, schatten69, hoeksema91}. The field
is most easily determined from the image-dipole formula derived by
\citet{ska07}.  For a dipole with moment $\vec{d}$ located at
a point $\vec{r}_d$
inside the Sun, and a source surface at radius $R_S$, the magnetic
field $\vec{B}$ is determined from the potential $\Phi$
via $\vec{B} = - \nabla \Phi$, where $\Phi$ is given by:
\begin{equation}
\Phi = \frac{\vec{d} \cdot (\vec{r} - \vec{r}_d)}{| \vec{r} - \vec{r}_d|^3}
       - \frac{R_S r_d^3 \vec{d} \cdot (R_S^2 \vec{r} - r^2 \vec{r}_d)}
              {| r_d^2 \vec{r} - R_S^2 \vec{r}_d|^3}.
\end{equation}
This field satisfies the source-surface boundary condition
that $B_\theta=B_\phi=0$ at $r=R_S$, since $\Phi=0$ there.
The advantage of this formulation is that most active regions can be
approximated by a collection of dipoles, and one can build up a
field of arbitrary complexity by simply adding a series of dipoles of the
form of Eq.~(1).  Each dipole is specified in terms of
its position in spherical
coordinates $\vec{r}_d=r_d\vec{\hat r}(\theta_d,\phi_d)$,
where $r_d$, $\theta_d$, and $\phi_d$ specify the location of
the dipole, and the spherical components
of its dipole moment, $\vec{d}=(d_r,d_\theta,d_\phi)$.

Figure~2 shows the field computed from Eq.~(1) for the case of
two dipoles: a sun-centered global dipole with a dipole moment
of unit magnitude directed along the north polar axis,
and an equatorial ``active region''
dipole at $\vec{r}_d=0.8R_\odot\vec{\hat r}(90^\circ,0^\circ)$
with a northward-pointing dipole moment $\vec{d} = (0,-0.2,0)$.
The source surface radius is chosen as $R_S=4R_\odot$, though
the exact value is not critical for our argument.
Note that for
convenience in viewing the magnetic field, we have selected the dipole
parameters so that the system has symmetry across both the equatorial
$(\theta = 90^{\circ})$ and meridional $(\phi = 0)$ planes. Also, for ease of
viewing, we show in the Fig.~2 only the front hemisphere defined by
the angular region $(15^{\circ} \le \theta \le 90^{\circ})$ and
$(-90^{\circ} \le \phi \le 90^{\circ})$.

The solar surface, the photosphere, corresponds to the gray grid in
Fig.~2. The colored contours on this
surface correspond to contours of radial flux, indicating the presence
of the active region dipole at the equator. We selected the
parameters for the active region dipole so that its structure would be
easily resolved.  It is evident from Fig.~2 that the region is
large compared to real active regions, which are generally only a few
degrees in angular extent. On the other hand, the maximum field
strength at the dipole center is only $\sim 20$ times that of the polar
region, which is much less than the corresponding ratio for solar
active regions, so the flux ratio between the active region and global
background field is approximately correct. This ratio is the important
parameter to obtain a coronal hole extension.

The thick black line along the equator is the $B_r = 0$ contour,
i.e., the polarity inversion line.  The thick black line above the
solar surface is the polarity inversion line at the source surface,
i.e., the bottom of the HCS. Red field lines are traced at equal
intervals along the HCS down to the solar surface.  These define the
boundary between open and closed field lines. As expected, the effect of the
equatorial dipole is to pull the open-closed boundary down to lower
latitudes; in other words, to create a low-latitude extension
of the coronal hole, which can be seen as the gray shaded
region in the Figure. Far
from the dipole, the coronal hole boundary is at a latitude of
$\sim 54^\circ$, whereas at the meridional symmetry plane the boundary
drops down to $\sim 26^\circ$.

For the large spatial scale of our active region dipole, the
extension of the coronal hole down to low latitudes is gradual rather
than in the form of a distinct ``elephant trunk'', but the basic
effect is clearly present. There is no open-field corridor in Fig.~2,
but let us now add another dipole to the system, displaced
$20^{\circ}$ in both latitude and longitude from the equatorial one
and a factor of five times weaker.  This dipole is located at
$\vec{r}_d=0.8R_\odot\vec{\hat r}(70^\circ,20^\circ)$
with a primarily southward-pointing dipole moment $\vec{d} = (0,0.05,0)$.
In order to maintain the equatorial and meridional
symmetry, as mentioned earlier, we actually add 4 dipoles
symmetrically located about the equatorial and meridional planes.

The resulting field is shown in Figure~3. The effect of the additional
dipoles is to add high-latitude polarity inversion lines to the
system. These ``squeeze'' the open-flux extension of Fig.~2 to form a
narrow corridor and a low-latitude coronal hole.  As in Fig.~2, red
field lines are traced from equidistant footpoints along the HCS down
to the solar surface. The red footpoints at the photosphere appear to
traverse the boundary of the low-latitude hole and then jump abruptly
to the polar hole boundary, which implies that the mapping defined by
the field develops extreme gradients in the region connecting the two
holes. To clarify this point, we have traced two sets of field lines,
colored in blue, from footpoints that are closely located at the HCS. The
corresponding solar footpoints are much more widely spaced, running
along the corridor. The resulting structure, Fig.~3, looks very
similar to the mapping drawn in Fig.~1, in that the closely spaced
pairs of points a$^\prime$,b$^\prime$ and c$^\prime$,d$^\prime$ at the
HCS map to far-separated points a,b and c,d at the solar surface. Note
also that although the footpoints of the two sets of blue lines
approach each other very closely at the photosphere, they are far
separated at the HCS, by a distance of order $R_\odot$. This result
indicates that even though the low-latitude coronal hole has small area, it
contains a substantial magnetic flux. As is evident from the colored
contours in Fig.~3, the photospheric field strength
in the low-latitude hole is large due to the presence of the active
region dipole. 

The analytic model underlying Fig.~3 has similar topology
to the case shown schematically in Fig.~1.
The low-latitude coronal hole extension in Fig.~3 is connected
to the main polar hole by a corridor that becomes very narrow.
Furthermore, this type of
topology is not difficult to obtain. It is often observed in
quasi-steady MHD solutions for observed photospheric fields, as will be
shown below.  A similar corridor was found for Carrington rotation 1922
\citep{ska07}.

The question now is whether the open flux in the corridor connects to
large latitudes in the heliosphere. To answer this question, we trace
field lines from a set of photospheric footpoints lying on a
latitudinal line segment spanning the narrowest width of the corridor,
which is only of order $5{,}000\,{\rm km}$ at the photosphere.
Fig.~4b shows the
footpoints and the field lines (green) near the photosphere and Fig.~4a
shows where they map to on the source surface. We note that the
corridor maps to high latitudes.  In fact, for this analytic case, the
corridor mapping defines a QSL arc that reaches latitudes $> 45^{\circ}$,
greater than that of the observed slow wind.

This result, that the corridor maps to heliospheric latitudes far
above the HCS, is robust in that it is not sensitive to the exact
position of the secondary dipole. The position and geometry of the
corridor, on the other hand, is very sensitive to the photospheric
flux distribution. For example, its width would change or even become
singular \citep{titov11}, and its location would change substantially
if the secondary dipoles were moved in longitude.  Based on flux
conservation arguments, and the fact that the heliospheric magnetic
field is almost uniform in latitude, we can argue that the angular
extent of the QSL arc, however, would be expected to depend primarily
on the ratio of the flux in the low-latitude coronal hole extension to
that in the polar hole.  For example, in the extreme case that the
fluxes were equal, the corridor mapping would be expected to reach the
heliospheric pole ($90^\circ$ from the HCS!), irrespective of the
geometry of the corridor or of the coronal holes.

\section{The S-Web Model}

If the width of the corridor at the photosphere is small compared to
the scale of typical motions there, such as the supergranular flow, we
expect that the whole corridor will continuously disrupt and reform at
the photosphere and, consequently, closed-field plasma will be
released by reconnection all along the QSL arc in the heliosphere.
Therefore, the topology of Fig.~2 may be able to resolve the slow wind
paradox. The overriding question, however, is whether there are enough
such corridors and corresponding QSL arcs in the heliosphere to
account for the slow wind that is observed. The flux distribution of
Fig.~2 produces only one such arc, which would certainly not be
sufficient to reproduce the observed slow wind. There are two issues
that must be addressed, the number of arcs (their density and extent
on the Sun and heliosphere), and the amount of mass and energy that
each arc can be expected to release. In this paper we concentrate on
the first issue and only briefly discuss the second in Section 4
below, because addressing this issue requires fully dynamic
calculations.

In order to address the issue of the number of QSL arcs, we calculated
the quasi-steady model for an observed photospheric flux
distribution. Figure~5a shows the photospheric radial field as derived
from MDI observations on SOHO \citep{scherrer95} for a time period
preceding the August 1, 2008 total solar eclipse. This calculation was
used to predict the structure of the corona prior to the eclipse,
using magnetic field data measured during the period June 25--July 21,
2008.  The prediction compares very favorably with images of the
corona taken during the eclipse in Mongolia \citep{rusin10}.  Note
that the high resolution of the calculation captures the details of
many small-scale bipoles in the photospheric magnetic field
\citep{harvey85}.  This has been incorporated into the idea of the
``magnetic carpet'' \citep{schrijver97}.  We also show the polarity
inversion line $B_r=0$ slightly above the photosphere, at
$r=1.05R_\odot$ to delineate the magnetic polarity of the large-scale
structures.  (The polarity inversion line in the photosphere itself
shows an enormous complexity that overshadows its usefulness to
discern the large-scale magnetic polarity.)

The quasi-steady model was calculated by using the 3D MHD code MAS.
The MAS code and its implementation are described in detail by
\citet{mikic94}, \citet{mikic99}, \citet{linker99}, and
\citet{lionello09}.  MAS solves the time-dependent MHD equations,
including a realistic energy equation with optically thin radiation
and thermal conduction parallel to the magnetic field.  Given the
magnetic field at the photosphere and an assumption for the coronal
heating source, the MHD equations are advanced until the magnetic
field settles down close to steady state.  MHD models are generally
considered to be the most sophisticated implementation of Parker's
solar wind theory because they incorporate all the essential physics,
including the balance between gas pressure and Lorentz force.  An
important assumption is the form of coronal heating, which is
prescribed empirically at the present time since the coronal heating
process is still unknown.  The parameters of the empirical heating
model are constrained by observations of coronal emission in EUV and
X-rays \citep[e.g.,][]{lionello09}, as well as by solar wind
measurements. Details on the assumed form for the heating and on
the thermodynamics used in the MAS code can be found in
\citet{mikic07} and \citet{lionello09}.

In order to capture as much of the photospheric magnetic structure as
possible, we ran the MAS code with unprecedented resolution. Our
calculation used more than 16 million mesh cells and was run on over
4000 processors of NSF's Ranger supercomputer at the Texas Advanced
Computing Center, making it possible to include much of the
small-scale structure of the photospheric field in both the quiet sun
and in coronal holes, as shown in Fig.~5a.  These calculations are
unique in the degree to which they capture the small-scale structure
of the measured magnetic field.

Figure~5b shows the distribution of open and closed magnetic field
regions at the solar surface as determined by the model.
It is evident that there are many
low-latitude coronal hole extensions, similar to that in Fig.~3, but
with much more structure.  Several of these extensions appear to be
disconnected from the main polar holes,
but this is partly due to the limited resolution of the figure.  A few of
these coronal hole extensions are indeed connected by very thin corridors
in the photosphere, though many are only linked to the polar
coronal holes in a singular manner, as described in detail by \citet{titov11},
and as discussed further below.

The open field pattern in Fig.~5b is clearly complex, but the
important issue is the degree of complexity of the mapping into the
heliosphere and, in particular, the structure of the separatrices and
QSLs there.  We determined the open field mapping in great detail by
tracing tens of millions of magnetic field lines.  The topology of
this mapping, as evidenced by structures such as separatrices and
QSLs, is most easily seen by analyzing the squashing factor $Q$
\citep{titov02,titov07}. $Q$ is a measure of the distortion in the
magnetic field mapping, and is directly related to the gradients in
the connectivity. QSLs are regions of very large $Q$; we generally
define them as any region with $Q > 10^3$.  True separatrices such as
the HCS have infinite $Q$, because the mapping is singular there, but
when computed numerically they appear as surfaces with very large
(unresolved) values of $Q$.  The gray arc at $r=5R_\odot$ in Fig.~1 is
a QSL in the open field, and consequently would be a region of high
$Q$.  The green HCS would also be a region of high (infinite) $Q$. As
will be seen below, a high-resolution analysis of the $Q$ properties
of our MHD simulation is extremely informative.

Figure~6a shows $Q$ in a meridional plane at a central Carrington
longitude of $23.33^\circ$ at the time of the eclipse at 10:21UT,
while Figure~6b shows magnetic field lines traced from the vicinity of
the solar limbs at the same time.
We see that $Q$ outlines the boundary between open and closed field,
which is a true separatrix surface, but it is apparent that
there is much more detailed structure in both the closed and
open field regions.
The complex structure of $Q$ in the closed-field region is expected;
it simply reflects the fact that the photospheric field consists
of many small bipoles; but, there is also substantial structure
in the open field near the open-closed boundary.
Note the presence of a ``pseudostreamer'' on the NE limb, a feature
that has been discussed by \citet{wang07}. The relationship of
pseudostreamers to open hole corridors and the S-web is discussed in
detail in \citet{titov11}

Figure 7a shows $Q$ in the spherical surface at $r=10R_\odot$
using a logarithmic scale.  This is the structure that is
expected to map into the inner heliosphere (appropriately wrapped
into a spiral magnetic field by solar rotation), since the magnetic
field has reached its asymptotic structure by this radius.
The thick black line is the heliospheric current sheet (at
which $B_r$ reverses sign).
Figure~7b shows the magnitude of $B_r$ at the same radial surface
$r=10R_\odot$. Note that the choice of $10R_\odot$ is not crucial.  
Any surface in the heliosphere (where the field is all open)
yields similar results.

It is important to emphasize that the apparent structure in $Q$ expresses
only the connectivity of the open field, not its actual
magnitude.  
In spite of the enormous magnetic
complexity at the solar surface, the radial field distribution in the
heliosphere is completely unremarkable, Fig.~7b. There is a single polarity
inversion line denoting a single HCS, as is generally observed
near solar minimum, and
this HCS runs more or less equatorial. The radial field is essentially
uniform away from the HCS, as would be expected from simple pressure
balance.  (Careful examination of the plot of $B_r$ shows that there
is a faint semblance of the structure that can be seen in $Q$,
but it is only a small perturbation.) 

On the other hand, the $Q$ map at this surface {\it is} remarkable,
indeed, Fig.~7a.  We see that surrounding the HCS is a broad web of
separatrices and QSLs of enormous complexity.  There are at least
four striking features of this S-web. First, it has an angular extent
in latitude of approximately $40^\circ$, sufficient to account for
the observed extent of the slow wind. Note also that the angular
extent does vary with longitude, but only by a factor of two or so.
Second, the HCS is not necessarily in the center of the S-web, but is
sometimes near its edge. This can explain the frequent observation
that the HCS is usually not centrally located within slow wind streams
\citep[e.g.,][]{burlaga02}. Third, the boundary between the S-web
layer and the featureless polar hole region is sharp; it is narrow
compared to the width of the S-web. This can explain the observation
that the transition from slow to fast wind as measured by the
composition data is narrow compared to the slow wind region itself
\citep{zurbuchen99}.

In order to explore the details of how coronal hole extensions connect
to the polar holes, we calculated coronal hole areas at different
heights in the corona.  Figure~8 shows the location of a region near
longitude $75^\circ$ and latitude $15^\circ$N in which we explored the
connection between the low-latitude coronal hole extensions (of
negative polarity, shown in blue) in detail.  It is evident that the
coronal hole extensions in this region appear disconnected from the
north polar hole in the photosphere, but connect with it low in the
corona (at heights approximately between $0.01R_\odot$ and
$0.02R_\odot$ above the photosphere).  Figure~9 shows explicitly how
these coronal holes connect in the low corona.  The three-dimensional
shape of the coronal hole boundary is shown as a green
semi-transparent surface in the low corona in the region detailed in
Figure~8.  This is the boundary between open and closed field regions.
The regions marked by A, B, and C show examples in which the
extensions of coronal holes are not connected in the photosphere, at
least by any measurable open-field corridor, but appear to connect
above the photosphere in the low corona.  These regions are also
indicated in Figure~8 for ease of cross-reference.  Despite the fact
that these coronal holes are ``disconnected'' in the photosphere, they
always remain topologically {\it linked} in a singular manner with the
polar coronal hole, as discussed by \citet{titov11}.

Finally, note that the connections of the high-$Q$ lines between the
neighborhood of the HCS and the photosphere and low corona that were
postulated by the uniqueness conjecture \citep{ska07} are largely
present, even though the insight from these new high-resolution MHD
simulations has led us to generalize the uniqueness conjecture.  We
have found that, in general, coronal hole extensions are sometimes
connected to the polar holes in the photosphere via narrow corridors,
as originally postulated \citep{ska07}, but in other instances they
are disconnected in the photosphere, but remain topologically linked
to the polar holes \citep{titov11}.  In either case, these connections
are responsible for the formation of the S-web.  It should be
emphasized that in order to capture the intricate structure of these
connections, very high resolution models are required that can
incorporate some of the complexity of the photospheric magnetic carpet
fields. Given sufficient resolution, the S-web should appear as a
generic feature of all quasi-steady models, including the PFSS. In
fact, the PFSS models should be more effective than the MHD for
studying the complex topology of the S-web, because they allow for
much higher spatial resolution than is possible with an MHD code. On
the other hand, for quantitative comparison with observations, the MHD
models should be more effective, because they include the gas thermal
and kinetic pressure forces and Lorentz forces that we know are
present in the real corona.

\section{Discussion}

The major conclusion from our results is that the underlying premise
of the streamer top model is valid. The slow wind is expected to
originate from the release of closed-field plasma due to the dynamic
rearrangement of the open-closed field boundary. The key new addition
of our S-web model to this picture is that the inherent complexity of
the photospheric field leads to a network of narrowly connected and
disconnected coronal holes that nevertheless always remain linked.
This produces a separatrix web in the heliosphere that extends the
release of slow wind to regions that significantly depart from the
HCS.  Hence, our model accounts for both the observed composition and
the broad extent of the slow wind.

One immediate prediction from the model is that the angular width of
the slow wind is determined primarily by the complexity of the flux
distribution in the photosphere.  This complexity produces a very
convoluted polarity inversion line in the low corona and an intricate
coronal hole pattern (Figure~5).  Our ability to identify the S-web
and its manifestations rests on high-resolution calculations that are
beginning to capture the multitude of small dipoles in the
photospheric magnetic field.  If the solar field were a pure dipole,
producing an inversion line that runs straight along the equator, then
only the polar coronal holes would be present and there would be no
separatrix web in the heliosphere. For this ``basal'' (though
idealized) slow wind case, if we assume that the dynamic broadening of
the open-closed boundary at the Sun is of order the scale of a
supergranule, $\sim 30{,}000\,{\rm km}$, the angular extent of the
wind would be only of order $3^\circ$--$5^\circ$, and would be
centered about the HCS. Of course, the solar field is never a simple
dipole.

At the present time we do not know if the complexity seen in Figures
5--7 is typical, or whether it is particular to this late declining
phase of Cycle 23.  It should be noted that the present minimum
appears to be somewhat different than the previous few minima. In
particular, the polar field strength is significantly weaker
\citep[e.g.,][]{luhmann09}.

The S-web model predicts that for time periods during which extensions
of coronal holes away from the main polar holes are less prevalent
than in Cycle 23, the angular extent of the slow wind region would be
smaller. In fact, there is clear evidence from radio scintillation
data \citep{tokumaru10} and recent Ulysses solar wind measurements
that the Cycle 23 minimum has a substantially broader and more
structured slow wind region than that of the previous cycle.  Indeed,
during the previous minimum (circa 1996), equatorial coronal hole
extensions were less common than during the recent solar minimum.
Further high-resolution numerical calculations will be needed to
address this result.

Another prediction of the model is that the slow wind region is
actually a mixture of winds. It is evident from Fig.~7 that the
separatrix web is not space-filling. There are regions within the
broad S-web band where the wind emanates from the low-latitude coronal
hole extensions. These regions are likely to have large expansion
factor, so that the wind will be slow compared to the fast wind from
the polar regions, but its composition will be different than that of
closed-field plasma. Our model, therefore, naturally explains the
observed variability of the slow wind composition.

A key aspect of the S-web model that has yet to be calculated is the
dynamic release of closed-field plasma. Although our quasi-steady
calculations allow us to investigate the topology of the field, and to
identify the structure of the separatrix web in the heliosphere, they
do not actually produce a slow wind with closed-field composition. For
this we need fully dynamic simulations that include the driving due to
photospheric motions (e.g., resulting from differential rotation) and
flux emergence. Such simulations are now being performed in 3D
\citep[e.g.,][]{edmondson09, edmondson10, linker10} for simplified
photospheric flux distributions and driving flows. These simulations
do verify the basic idea of the S-web model that open-field corridors
will form and evolve in response to photospheric motions
\citep{edmondson09}. Higher resolution simulations will be needed,
however, to test the model in detail. On the other hand, it seems
unlikely that dynamic calculations with the degree of structure
present in Fig.~7 will be feasible in the near future. It is likely
that a definitive treatment of the slow wind will require the
development of a statistical theory of the dynamics of the S-web
model.

\acknowledgments

This work has been supported by the NASA TR\&T, SR\&T, and HTP 
Programs. The work has benefited greatly from the authors'
participation in the NASA TR\&T focused science team on the
solar-heliospheric magnetic field. SKA thanks J. Karpen for invaluable
scientific discussions and help with the graphics.


\clearpage

\begin{figure}
\plotone{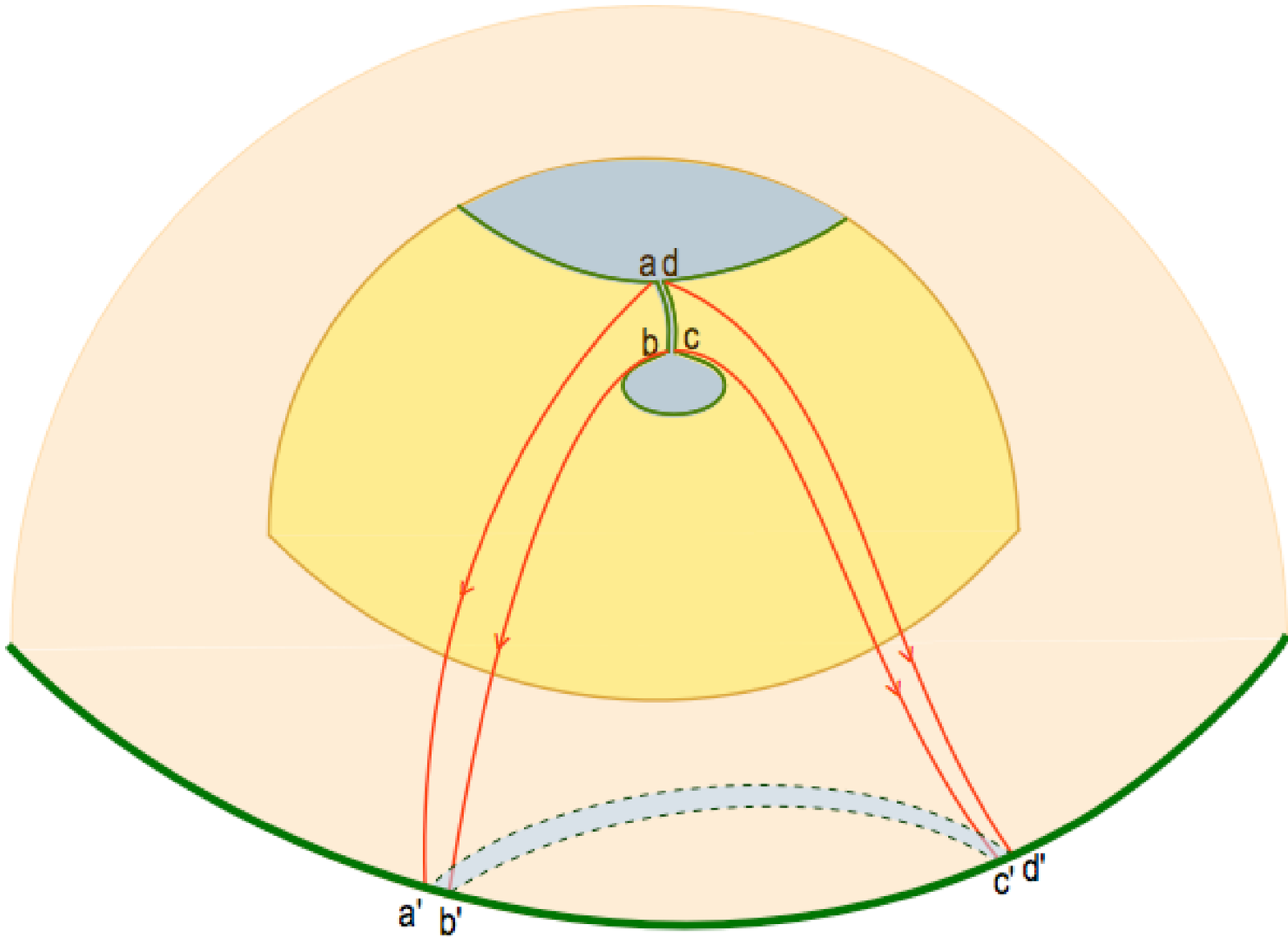}
\caption{ Magnetic field topology of an open field region consisting of
  a large polar coronal hole and a smaller low-latitude hole connected
  by an open-field corridor. The inner surface is the photosphere,
  with the dark gray and bright yellow regions corresponding to open
  and closed field respectively. The outer transparent surface is a
  radial surface in the heliosphere. The dark green line is the
  polarity inversion line and the light gray arc indicates where the
  open-field corridor maps to on this outer surface. 
\label{f1}}
\end{figure}

\begin{figure}
\epsscale{.75}
\plotone{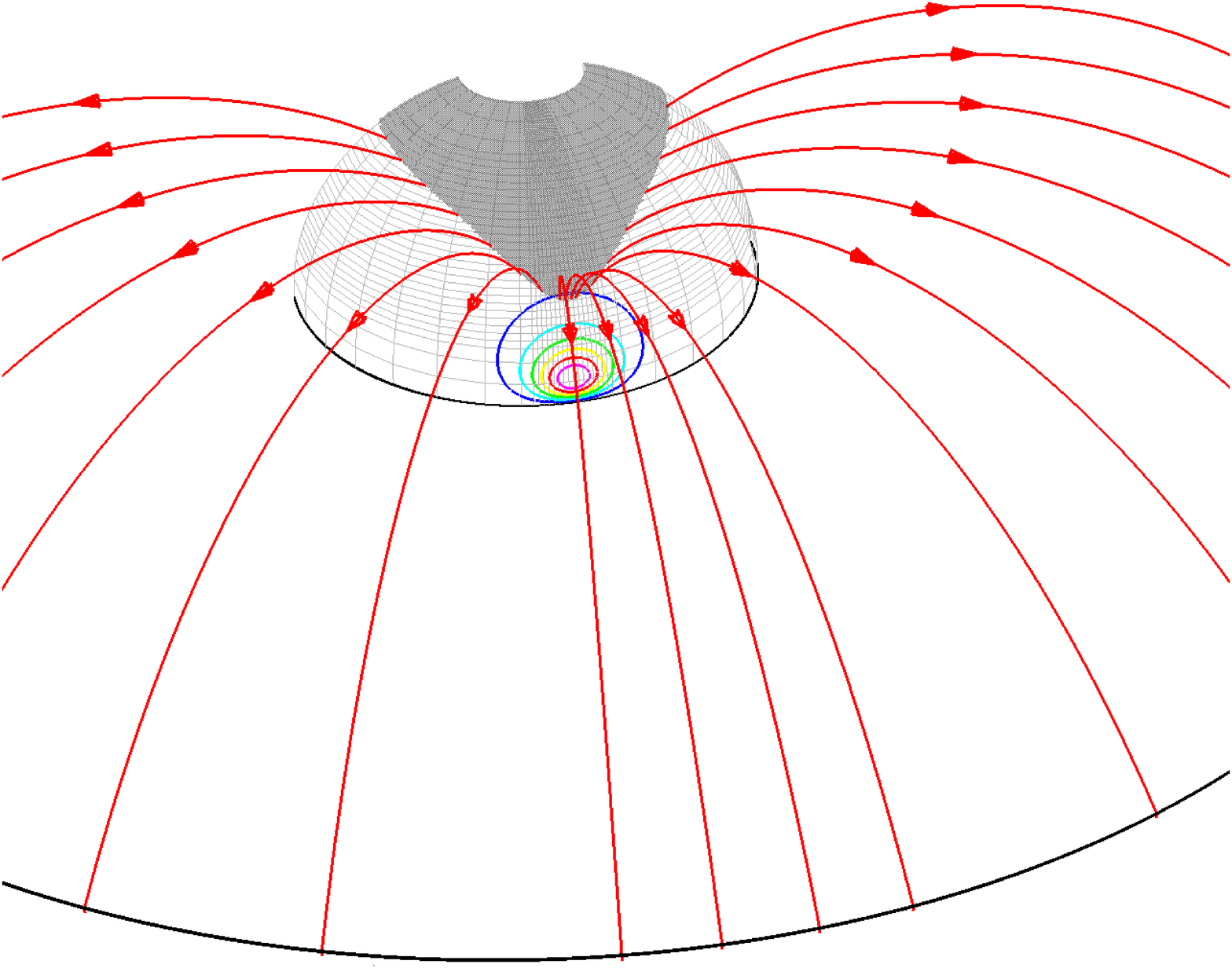}
\plotone{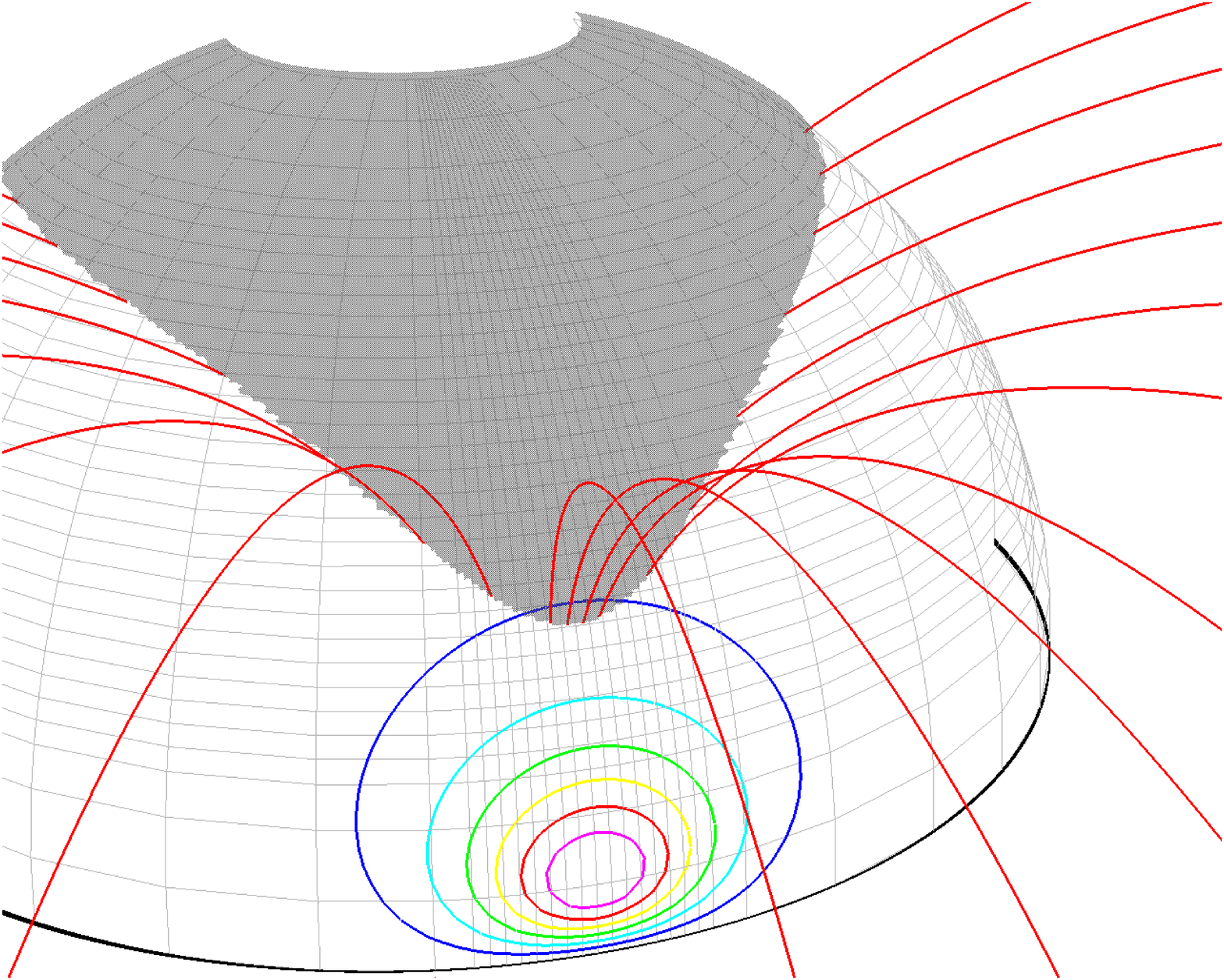}
\caption{ (Top) Open-closed magnetic field topology for a
  photospheric flux distribution due to a global dipole and an
  equatorial dipole. The gray shaded region indicates the polar
  coronal hole (the open flux region). The contours on the inner surface indicate radial
  field magnitude at the photosphere. The black lines correspond
  to the polarity inversion line at the photosphere and source
  surface. The red lines are magnetic field lines. 
  (Bottom) Close-up near the solar surface of the magnetic field above.
  \label{f2}}
\end{figure}

\begin{figure}
\plotone{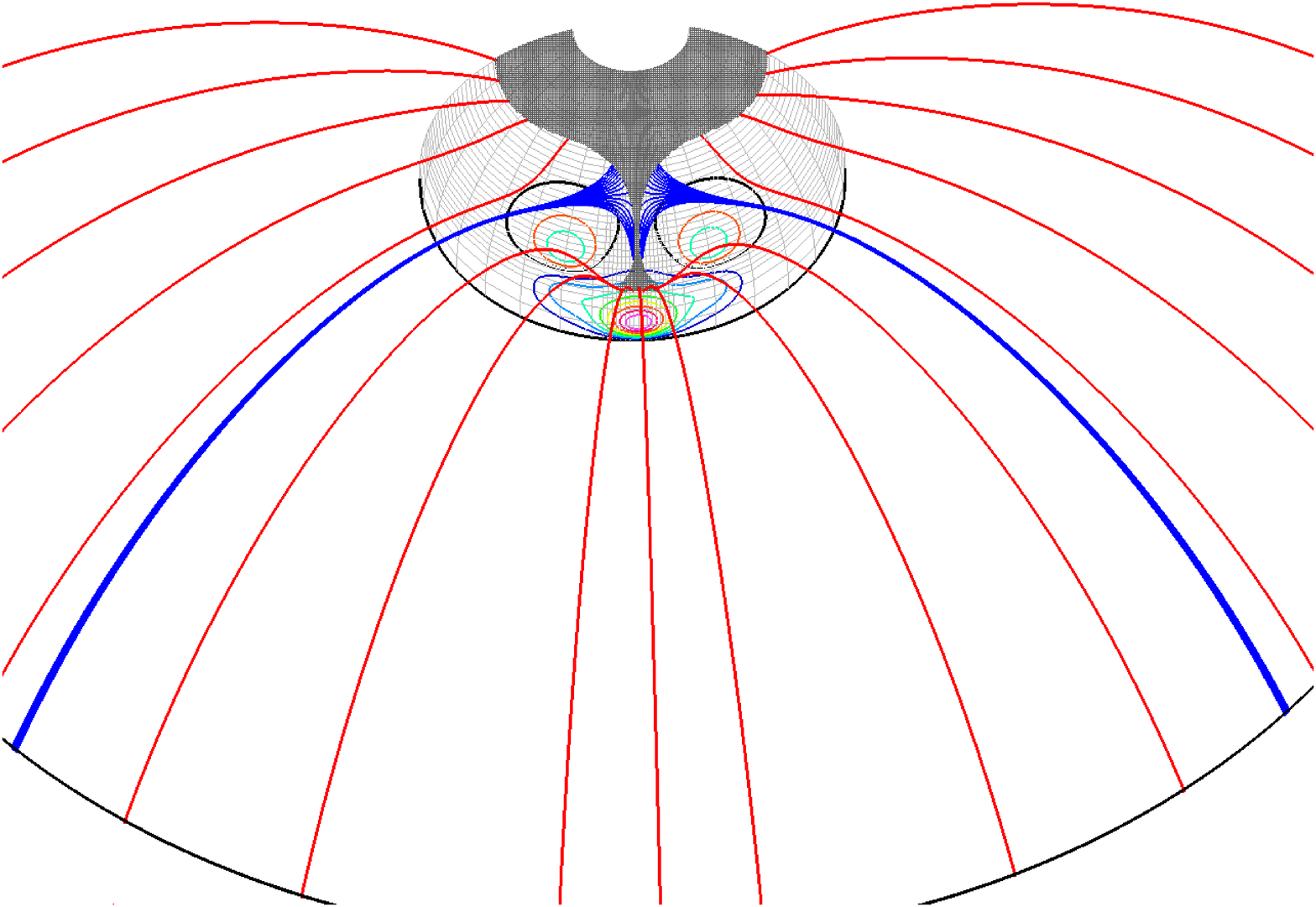}
\plotone{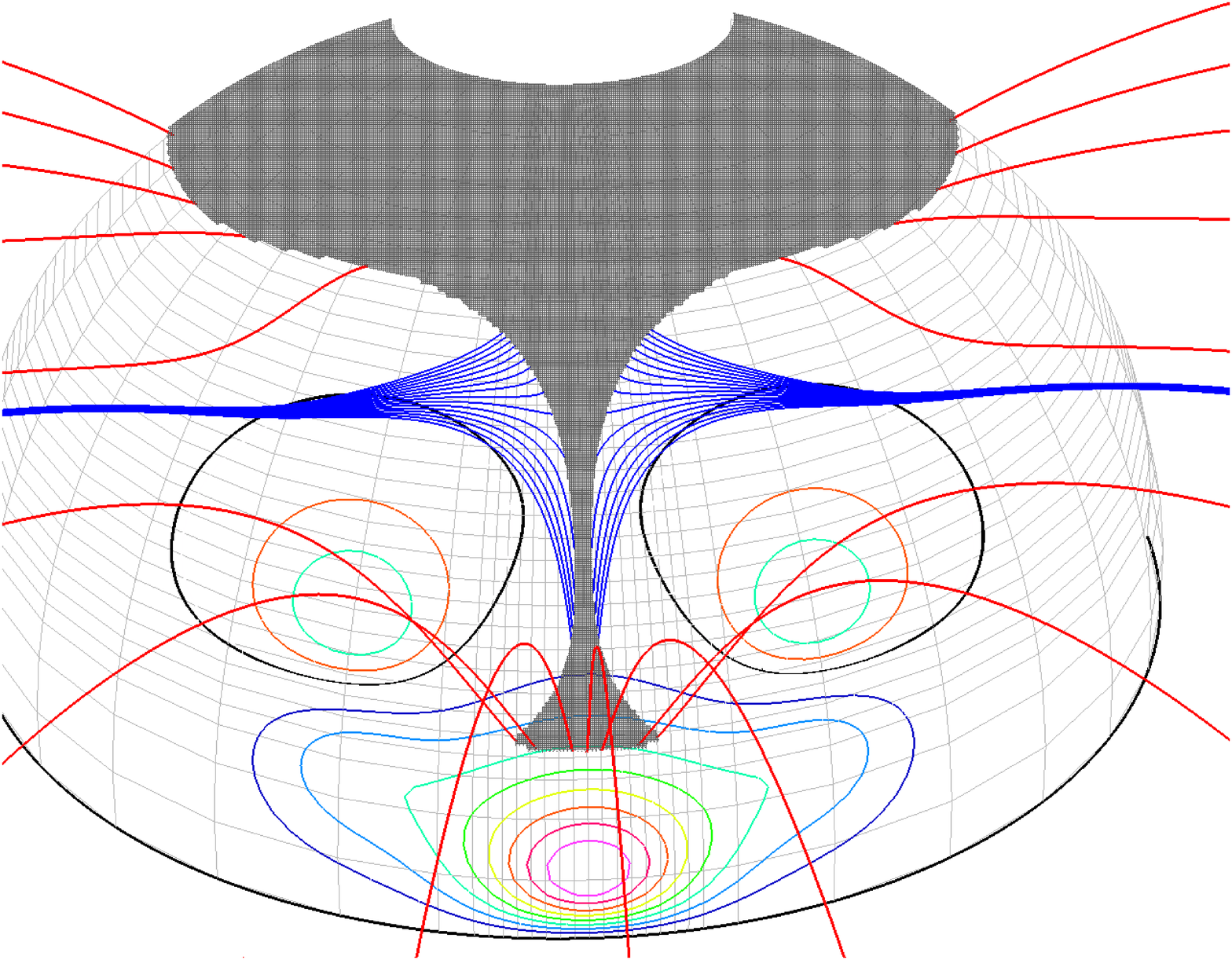}
\caption{ As in Figure~2, but for a flux distribution that includes
  additional high-latitude dipoles. Two additional polarity inversion lines can 
be seen at the photosphere. 
The blue field lines outline an
  open-field corridor. Note that the system is symmetric about the 
meridional plane $\phi = 0$.
  \label{f3}}
\end{figure}

\begin{figure}
\plotone{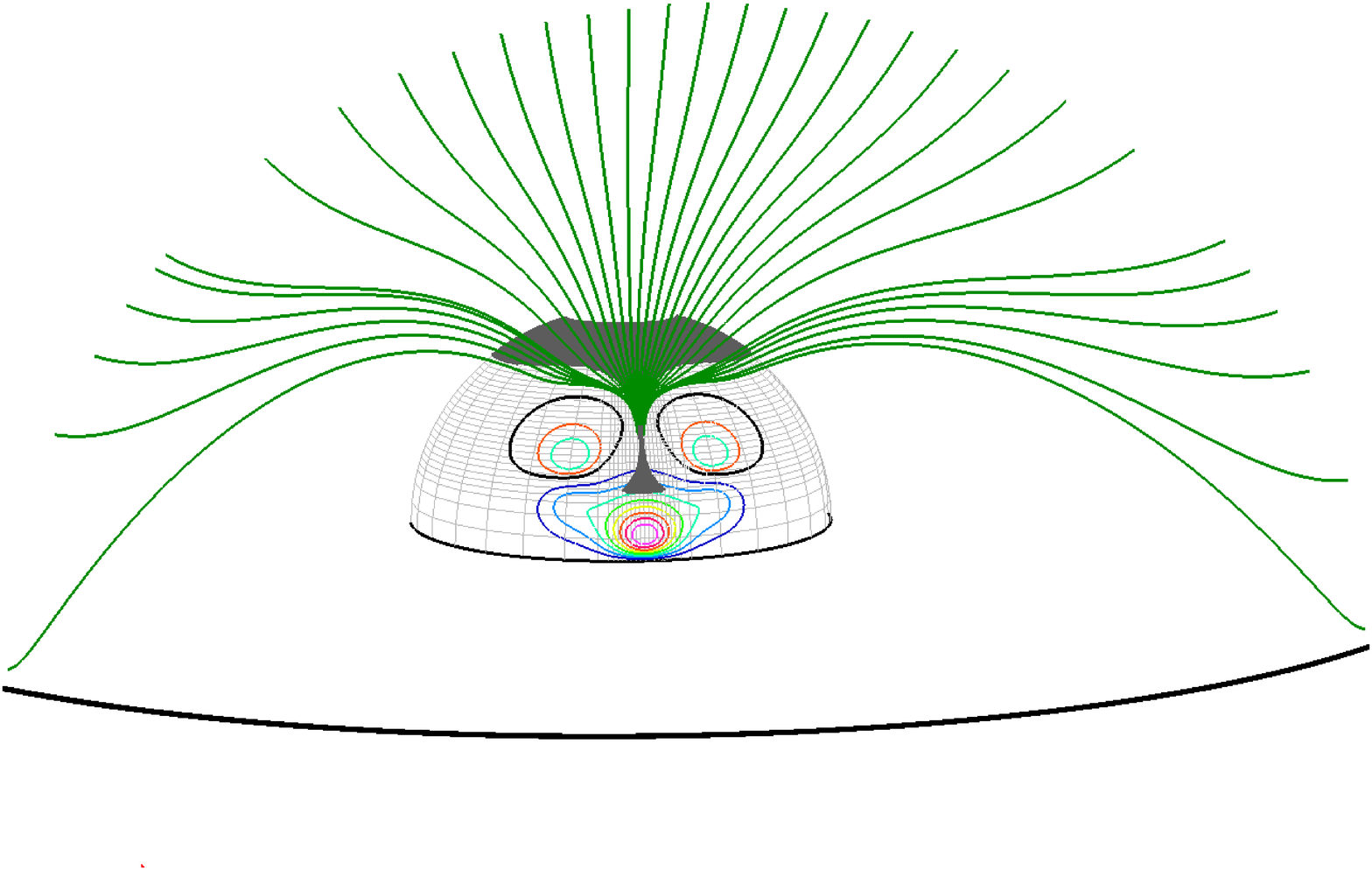}
\plotone{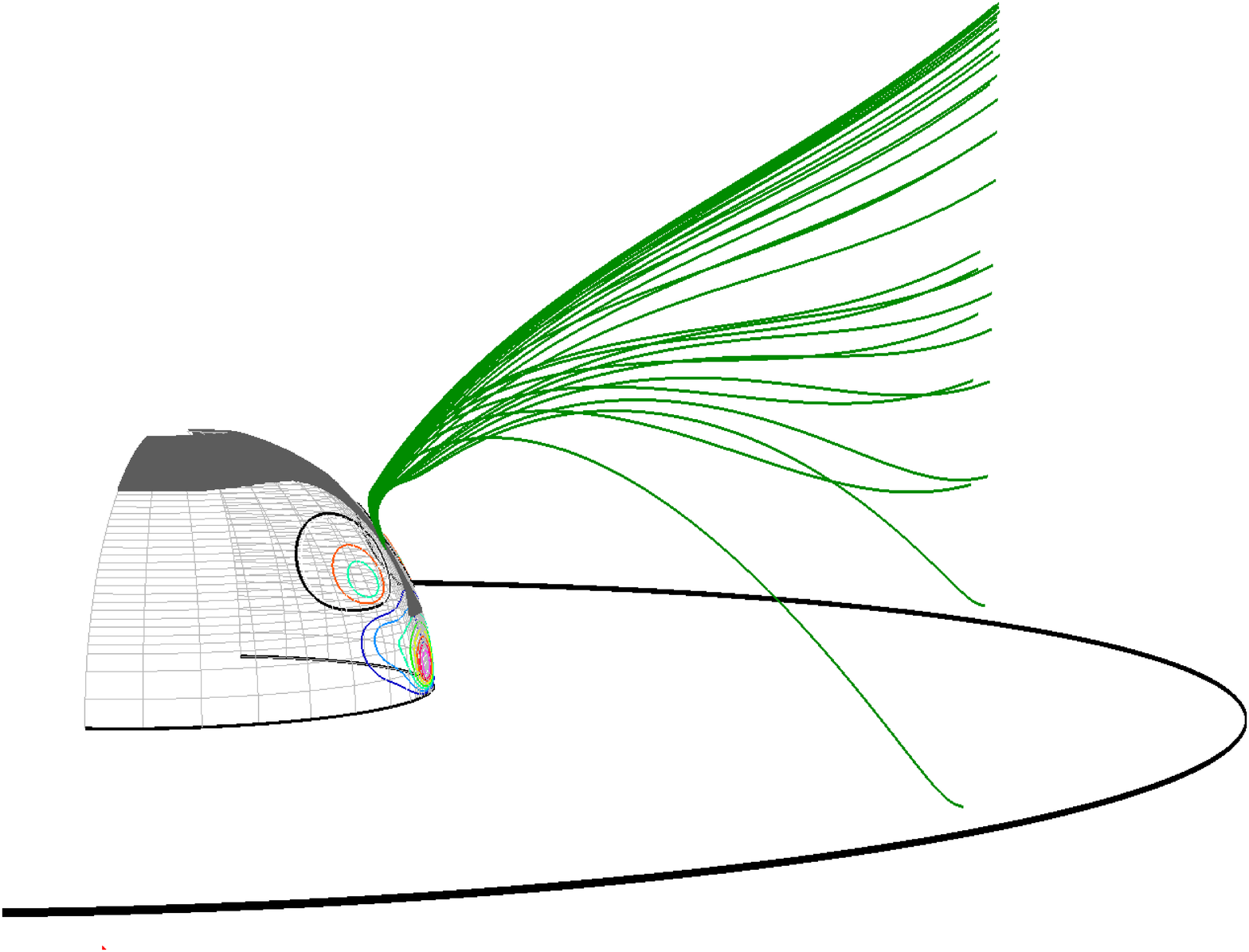}
\caption{ (Top) Open field lines (green) traced from photospheric footpoints 
along a line segment spanning the narrowest part of the corridor. 
The lines clearly extend to high latitude above the HCS. (Bottom) Close-up
near the solar surface showing the photospheric footpoints of the corridor
field lines. 
\label{f4}}
\end{figure}

\begin{figure}
\epsscale{1.}
\plotone{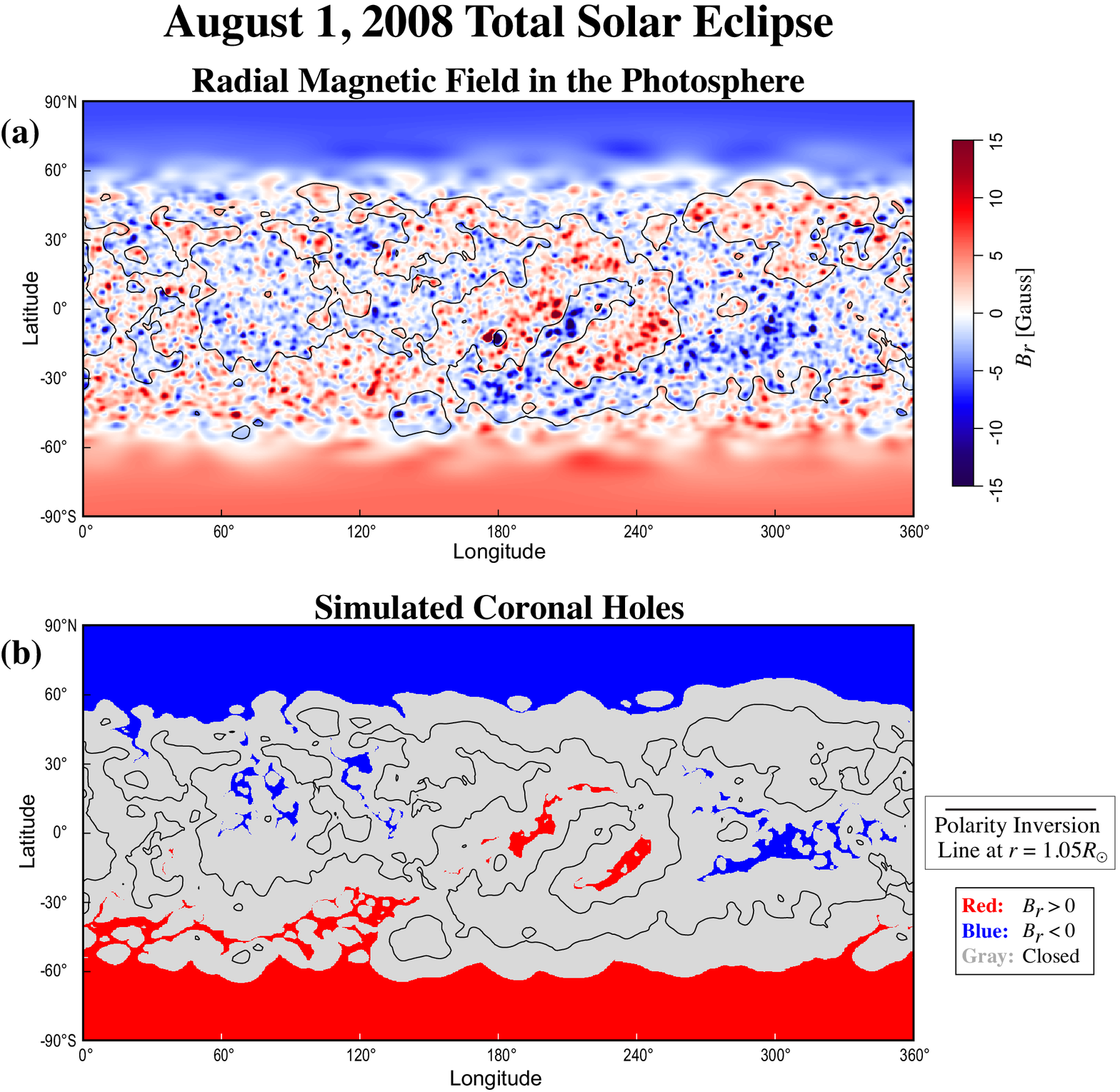}
\caption{ (a) Distribution of the radial component of the magnetic field
in the photosphere that was used in the MHD simulation to predict the
structure of the corona for the August 1, 2008 eclipse, as deduced from
MDI measurements.  (b) The open and closed field regions in the photosphere
as determined from the MHD solution.  The polarity
inversion line ($B_r=0$) at a height $r=1.05R_\odot$ is superimposed
on these images to aid in identifying the polarity of the large-scale
magnetic flux.
\label{f5}}
\end{figure}

\begin{figure}
\epsscale{1.}
\plotone{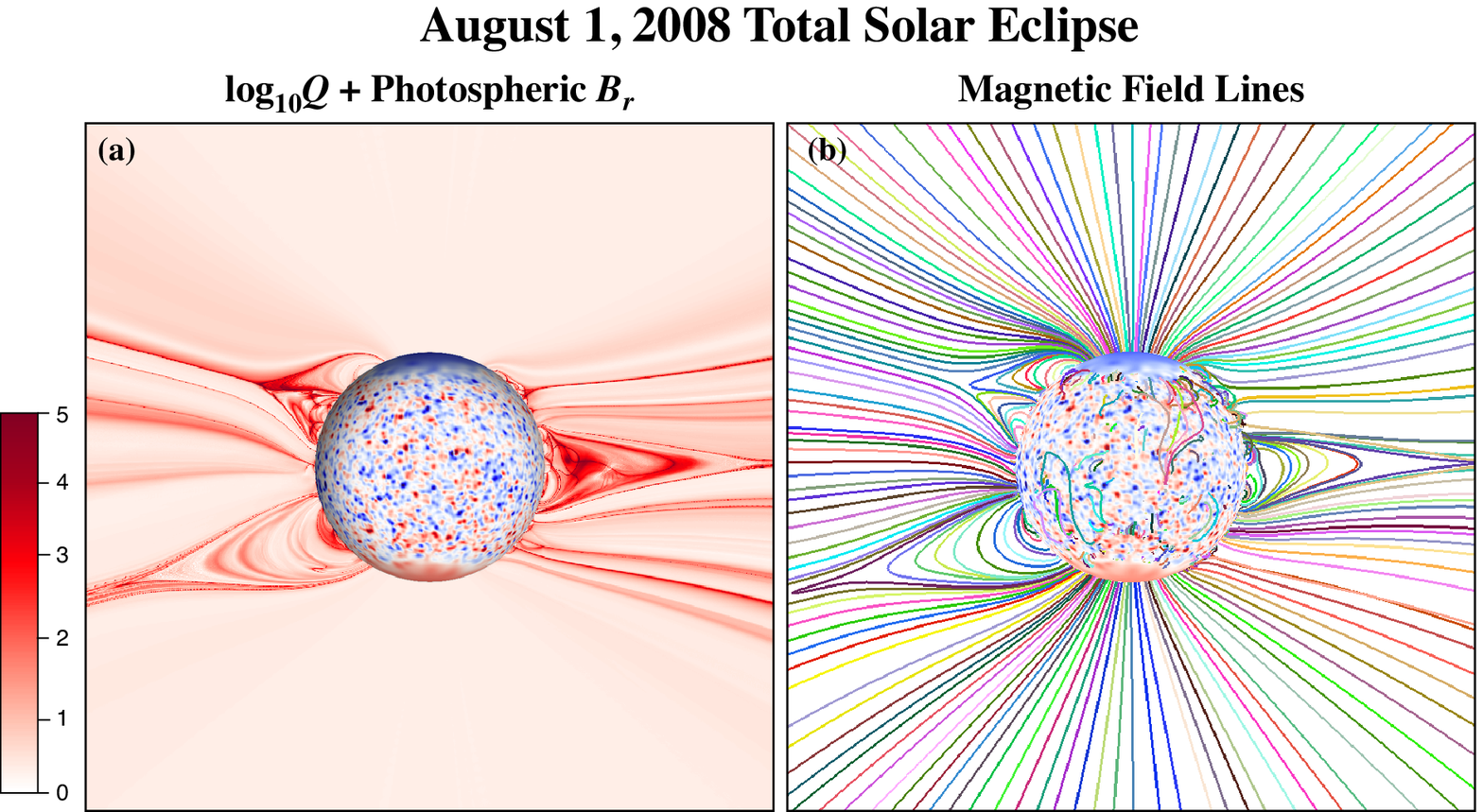}
\caption{ (a) Plot of the squashing factor $Q$ on a logarithmic scale
in a meridional plane at the time of the eclipse on August 1, 2008
at 10:21UT.  In this view, solar north is vertically up and
the $B_0$ angle is zero.  [At the time of the eclipse
$B_0=5.8^\circ$, so this view is slightly different than what would
have been observed.]  The Sun's surface is colored by the value
of $B_r$ with the same scaling as that in Fig.~5.
(b) Magnetic field lines traced from the vicinity of the limbs
at the same time, showing the structure of the open and closed
field regions.
\label{f6}}
\end{figure}

\begin{figure}
\epsscale{1.0}
\plotone{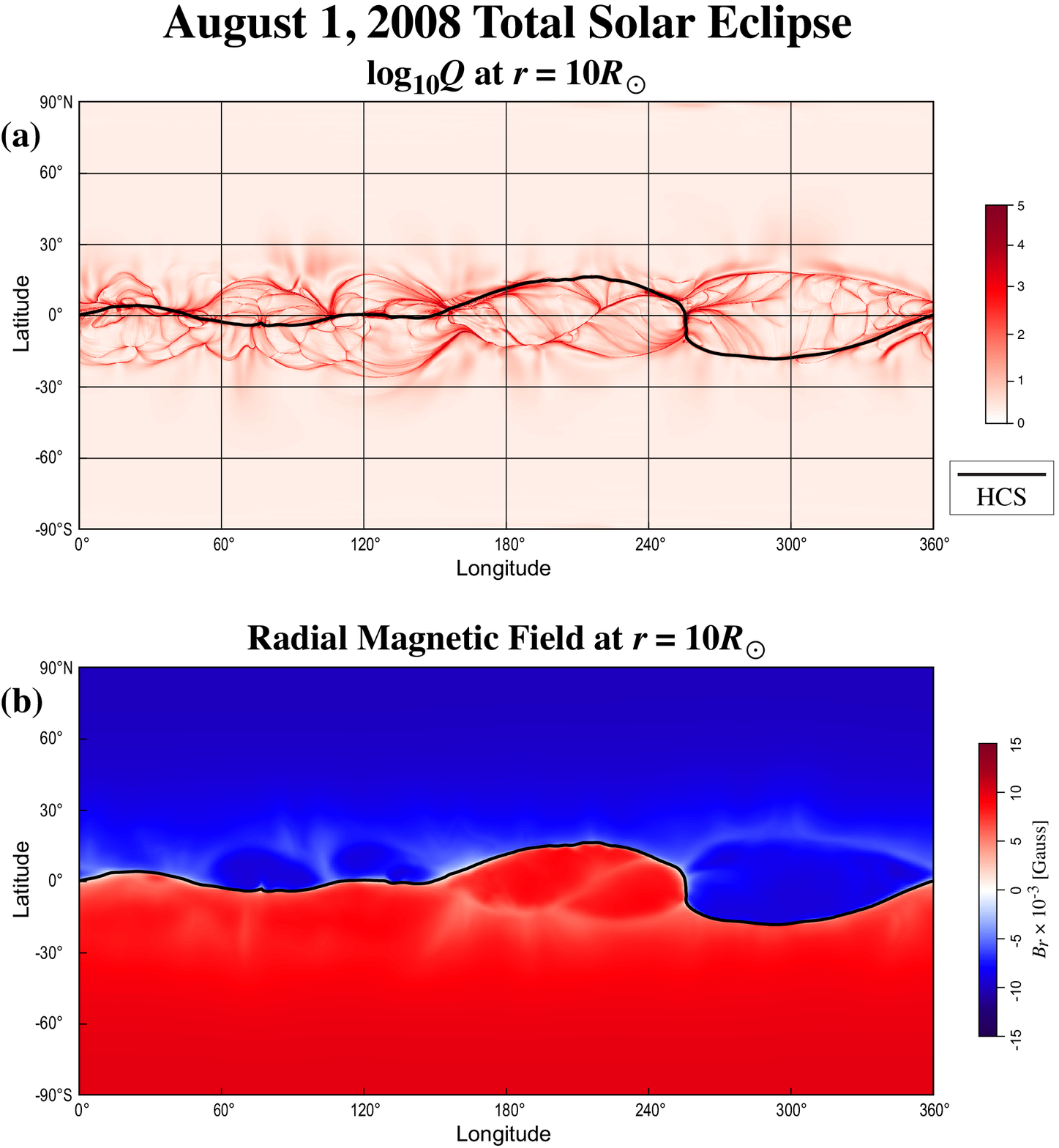}
\caption{ (a) Plot of the squashing factor $Q$ in the spherical
surface $r=10R_\odot$ on a logarithmic scale versus longitude
and latitude.  
(b) Plot of $B_r$ in the same spherical surface.  The HCS
(i.e., the location of $B_r=0$) is superimposed
on these images as a thick black line.
The complex structure in $Q$ in the vicinity of the HCS is produced
by the S-web.
\label{f7}}
\end{figure}

\begin{figure}
\epsscale{.9}
\plotone{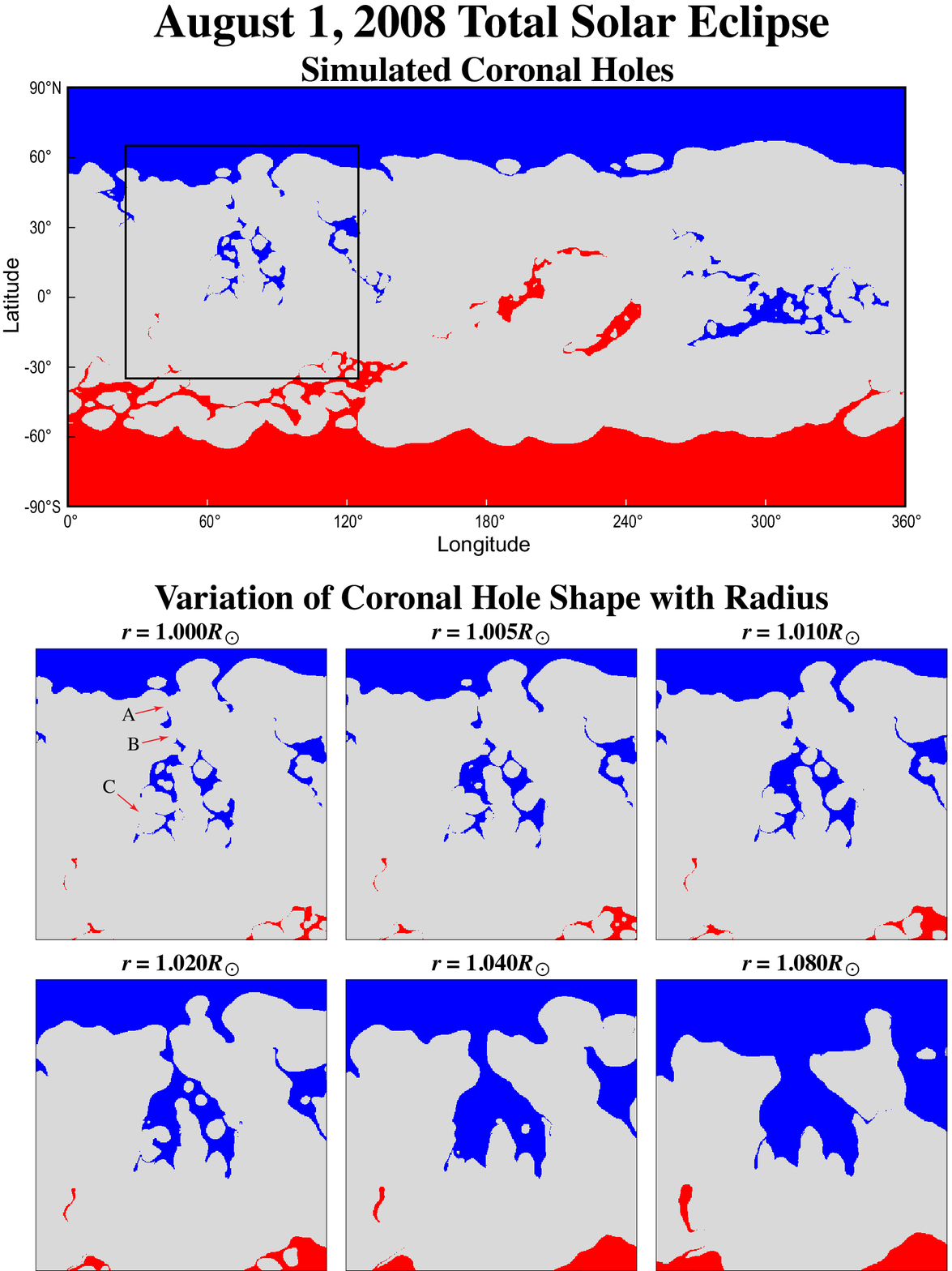}
\caption{ The variation of coronal hole shape with height above
the photosphere.  The top panel shows coronal holes at the
photosphere, as in Fig.~5b.
The black square shows a $100^\circ \times 100^\circ$ region
centered at longitude $75^\circ$ and latitude $15^\circ$N
that was used to compute the variation of
coronal hole shape with radius in the lower panels.
Note that the extended coronal holes connect to the polar
holes low in the corona.  The regions denoted by A, B, and
C are cross-referenced with the corresponding regions in Fig.~9.
\label{f8}}
\end{figure}

\begin{figure}
\epsscale{.9}
\plotone{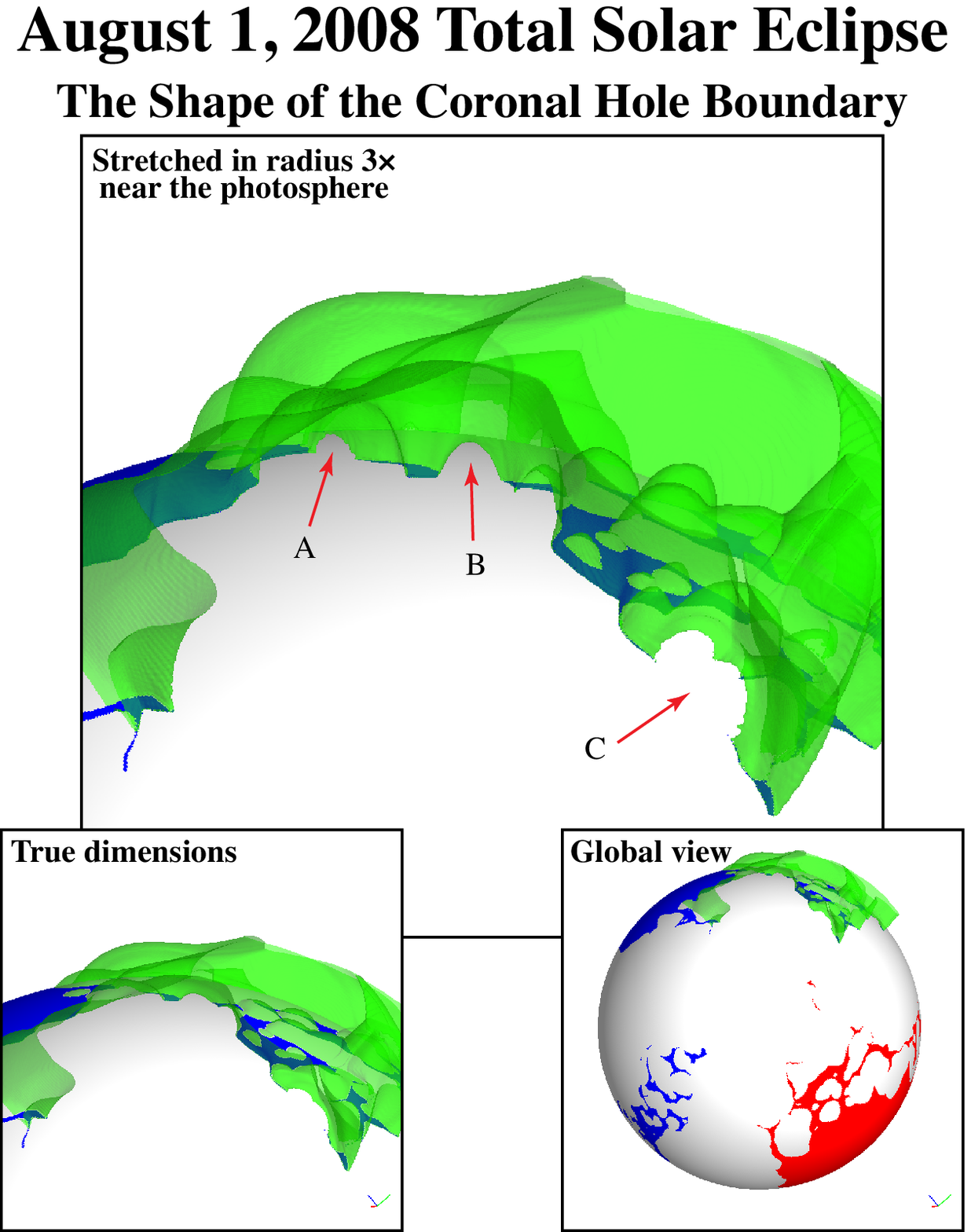}
\caption{ The three-dimensional shape of the coronal hole boundary
(semi-transparent surface) in the region detailed
in Figure~8, showing that some of the coronal hole extensions
(blue areas on the surface of the sphere) connect
with the north polar hole low in the corona.  The top panel shows a view
in which the surface is artificially stretched in radius
by a factor of 3$\times$ to show details near the photosphere.
The bottom left panel shows the same view without the radial stretching.
The bottom right panel shows the region detailed in the context of
the whole Sun.  The regions denoted by A, B, and
C are cross-referenced with the corresponding regions in Fig.~8.
\label{f9}}
\end{figure}

\end{document}